\documentclass[aps,pra,superscriptaddress,preprint,amsmath,amssymb,showpacs]{revtex4-2}
\usepackage[english]{babel}
\usepackage[utf8]{inputenc}
\usepackage{amssymb,bm}
\usepackage{graphicx}
\usepackage{amsmath}
\usepackage{color}
\usepackage{comment}
\usepackage[colorlinks=true,citecolor=blue,urlcolor=blue]{hyperref}
\usepackage{import}
\usepackage[T2A]{fontenc}
\usepackage{multirow}
\usepackage{caption2}
\usepackage{hyperref}
\usepackage{xspace}
\usepackage{xcolor}
\usepackage{appendix}
\usepackage{hyperref}

\begin{document}
	\begin{titlepage}
		
		\title{Relativistic effects in $\mbox{M1}$ radiative decays of heavy-light mesons}
		\author{A.~E.~Bondar}\email{A.E.Bondar@inp.nsk.su}
		\author{A.~I.~Milstein}\email{A.I.Milstein@inp.nsk.su}
		\affiliation{Budker Institute of Nuclear Physics of SB RAS, 630090 Novosibirsk, Russia}
		\affiliation{Novosibirsk State University, 630090 Novosibirsk, Russia}

		\date{\today}

\begin{abstract}
We discuss the $\mbox{M1}$ radiative transitions $D^*\rightarrow D\gamma$, $D_{s}^*\rightarrow D_s\gamma$, $B^*\rightarrow B\gamma$, and $B^*_{s}\rightarrow B_s\gamma$. A relativistic potential model is proposed.  The corresponding Hamiltonian, when expanded to terms of the order of $v^2/c^2$, where $v$ are the quark velocities, coincides with the Breit Hamiltonian. This model allows making predictions for the widths of radiative transitions of mesons with one light quark. Taking into account relativistic effects is especially important for the transitions $D_{s}^{*+}\rightarrow D_s^+\gamma$ and $D^{*+}\rightarrow D^+\gamma$, where there is a large compensation in the magnitude of the magnetic moment of these mesons. Our results are consistent with known experimental data.	
\end{abstract}

\maketitle
 \end{titlepage}

\section{Introduction}

The $D^*$ meson decays are dominated by the radiative $D^*\to D\gamma$ decays ($\gamma$ transitions) and the strong $D^*\to D\pi$ decays with pion emission. Due to the experimental difficulties in measuring the total widths of $D$ mesons, the data on the partial widths of these decays are incomplete. 

%The first measurement of the $D^*$ meson width was made by the CLEO collaboration for the charged $D^{*+}$ meson, $\Gamma(D^{*+})=(96\pm 4\pm 22)\,\textrm{KeV}$~\cite{CLEO:2001sxb} (the first error is statistical, the second is systematic). 

At the moment only the $D^{*+}$ meson width is known. The BABAR collaboration measured this width with best precision: $\Gamma(D^{*+})=(83.3\pm 1.2\pm 1.4)\,\textrm{KeV}$~\cite{BaBar:2013thi}.
Using the measured total and fractional widths of $D^{*+}$ and the relative decay rates of 
 $D^{*0}$~\cite{ParticleDataGroup:2022pth}, and also assuming isotopic invariance for $D\pi$ decays, one can estimate the radiative width of $D^{*0}$, $\Gamma(D^{*0}\to D^0\gamma)=20.2(4.5)\,\mbox{KeV}$.

Radiative decays of $D^*$ mesons can provide information on the dynamics of hadronic processes. They allow testing various approaches in QCD. There are many theoretical attempts to calculate the widths of these decays. The radiative decays of $D^*\to D\gamma$ have been studied in the framework of the QCD sum rules (QCDSR)~\cite{Eletsky:1984qs, Aliev:1994nq, Dosch:1995kw, Zhu:1996qy}, the light-cone sum rules (LCSR)~\cite{Aliev:1995zlh, Li:2020rcg}, and various quark models~\cite{ODonnell:1994qrt, Ivanov:1994ji, Deandrea:1998uz, Colangelo:1994jc, Deng:2013uca, Orsland:1998de, Jaus:1996np, Goity:2000dk, Ebert:2002xz, Choi:2007se, Cheung:2014cka}. Lattice QCD (LQCD) calculations were performed for the decays $D^*\to D\gamma$~\cite{Becirevic:2009xp} and $D^*_s\to D_s\gamma$~\cite{Donald:2013sra}. Calculations on the lattice~\cite{Donald:2013sra}  predict the decay width $\Gamma(D_s^*\to D_s\gamma)=0.066(26)\,\mbox{KeV}$.
This value disagrees with the light-cone sum rule result $\Gamma(D_s^*\to D_s\gamma)=0.18(1)\,\mbox{KeV}$~\cite{Choi:2007se} by $4\sigma$, and with the QCDSR result $\Gamma(D_s^*\to D_s\gamma)=0.25(8)\,\mbox{KeV}$~\cite{Aliev:1994nq} by $2\sigma$. The central value of this prediction
is an order of magnitude smaller than most other predictions available in the literature. Since $D^*_s\to D_s\gamma$ dominates the $D^*_s$ meson decay, it is important to consider this case more closely.

\begin{table}[ht]
	\caption{Decay widths of $\Gamma(D^*_{(s)}\to D_{(s)}\gamma)$ in $\mbox{KeV}$. The table is taken from Ref.~\cite{Tran:2023hrn}.}
	\label{tab:width_VPg}
	\begin{center}
		\begin{tabular}{c|c|c|c}
			\hline\hline
			Ref.  & $\Gamma(D^{*+}\to D^+\gamma)$  &  $\Gamma(D^{*0}\to D^0\gamma)$ & $\Gamma(D^{*+}_s\to D^+_s\gamma)$\\
			\hline
			CCQM~\cite{Tran:2023hrn}  &   $1.21(48)$    &    $18.4(7.4)$      &  $0.55(22)$\\
			LCSR (NLO)~\cite{Pullin:2021ebn}  & $0.96^{+0.58}_{-0.62}$ & $27.83^{+9.23}_{-9.50}$ & $2.36^{+1.49}_{-1.41}$ \\
			LQCD~\cite{Becirevic:2009xp}$^\dagger$,~\cite{Donald:2013sra}$^*$     & $0.8(7)^\dagger $ & $27(14)^\dagger $  &  $0.066(26)^*$ \\
			HQET$+$CQM~\cite{Cheung:2014cka}    & $0.9\pm 0.3$ & $22.7\pm 2.2$ & --  \\
			Bag Model~\cite{Orsland:1998de}  & 1.73 & 7.18 & --\\
				NJL Model~\cite{Deng:2013uca}
			 & 0.7 & 19.4 & 0.09
			 \\
			RQM~\cite{Colangelo:1994jc} & 0.46 & 23.05  & 0.38 \\
			RQM~\cite{Jaus:1996np}  & 0.56  & 21.69  & --\\
			RQM~\cite{Goity:2000dk}   & (0.94--1.42) & (26.0--32.0) & (0.2--0.3)\\
			RQM~\cite{Ebert:2002xz} & 1.04 & 11.5 & 0.19\\
			LFQM~\cite{Choi:2007se}  & $0.90(2)$ & $20.0(3)$ & $0.18(1)$\\		
			QCDSR~\cite{Zhu:1996qy} & $0.23(10)$ & $12.9(2.0)$ & $0.13(5)$\\
			LCSR~\cite{Aliev:1995zlh} & 1.50 & 14.40 & -- \\
			QCDSR~\cite{Aliev:1994nq} & $0.22(6)$ & $2.43(21)$ & $0.25(8)$\\				
			HQET$+$VMD~\cite{Colangelo:1993zq} & $0.51(57)$ & $16.0(10.8)$ & $0.24(24)$   \\
                NRQM~\cite{Close:2005se} & 1.8 & $32$ & $0.2$   \\
                Bethe-Salpeter~\cite{Jia:2024imm} & 0.84 & $19.4$ & $-$   \\
			Experiment~\cite{ParticleDataGroup:2022pth} & $1.33(36)$ & $20.2(4.5)$ & $0.114^{+0.066}_{-0.050}$ \\
			\hline\hline
		\end{tabular}
	\end{center}
\end{table}

Predictions for $\Gamma(D^*_s\to D_s\gamma)$ range from 0.07 $\mbox{KeV}$ to 2.4 $\mbox{KeV}$. It is interesting to estimate the width of the $D^*_s\to D_s\gamma$ decay using a recent measurement of the leptonic decay $D_s^{*+}\to e^+\nu_e$ made by the BESIII collaboration~\cite{BESIII:2023zjq}. This is the first measurement of the weak decay branching of a vector meson. The obtained value is $\mathcal{B}(D_s^{*+}\to e^+\nu_e)=(2.1^{+1.2}_{-0.9_{\textrm{stat.}}}\pm 0.2_{\textrm{syst.}})\times 10^{-5}$. Taking the branching ratio $\mathcal{B}(D_s^{*+}\to e^+\nu_e)/\mathcal{B}(D_s^{+}\to \mu^+\nu_\mu)$, one can obtain
\begin{equation}
\Gamma^{\textrm{total}}_{D_s^{*+}}=\frac{2.04\times 10^{-6}}{\mathcal{B}(D_s^{*+}\to e^+\nu_e)}\left(\frac{f_{D_s^{*+}}}{f_{D^+_s}}\right)^2\,\textrm{KeV}.
\end{equation}
Using the world average value $\mathcal{B}(D_s^{+}\to \mu^+\nu_\mu)$ and the result $\frac{f_{D_s^{*+}}}{f_{D^+_s}}=1.12(1)$ obtained from LQCD calculations, the BESIII collaboration found that the total width of $D_s^{*+}$ is $\Gamma^{\textrm{total}}_{D_s^{*+}}=0.122^{+0.070}_{-0.052}\pm 0.012\,\textrm{KeV}$~\cite{BESIII:2023zjq}, which is consistent with the LQCD prediction of $0.070(28)\,\textrm{KeV}$~\cite{Donald:2013sra}. The world average value is $\mathcal{B}(D^{*+}_s\to D^+_s\gamma)=0.936(4)$~\cite{ParticleDataGroup:2022pth}, which can be used to obtain $\Gamma(D^{*+}_s\to D^+_s\gamma)=0.114^{+0.066}_{-0.050}\,\textrm{KeV}$ and compared with the theoretical predictions given in Table~\ref{tab:width_VPg}.

In our work we discuss the $\mbox{M1}$ radiative transitions $D^*\rightarrow D\gamma$, $D_{s}^*\rightarrow D_s\gamma$, $B^*\rightarrow B\gamma$, and $B^*_{s}\rightarrow B_s\gamma$. We propose a relativistic potential model in which the Hamiltonian of the quark-antiquark interaction when expanded to terms of the order of $v^2/c^2$, where $v$ are the quark velocities, coincides with the Breit Hamiltonian, see, for example, Refs.~\cite{Pilkuhn:1979ps,Lee:2001wwa}. This model allows predictions to be made for the widths of radiative transitions of mesons with one light quark.

Our work is organized as follows. We first give a heuristic derivation of the effective Hamiltonian, which is a relativistic generalization of the potential quark model. This Hamiltonian contains not only the spin-dependent contributions describing the hyperfine splitting of mesons, but also the spin-independent contributions to the total energy of states. Using this Hamiltonian, we obtain formulas for the spectrum of states and the widths of radiative transitions. Next, we fix the parameters of the model based on experimental data and then obtain predictions for $D^*\rightarrow D\gamma$, $D_{s}^*\rightarrow D_s\gamma$, $B^*\rightarrow B\gamma$, and $B^*_{s}\rightarrow B_s\gamma$. In addition, we obtain quantitative predictions for the magnitude of the hyperfine splitting in the $b\bar s$ system and compare them with recently available experimental data~\cite{CMS:2025kat}. The conclusion  formulates the main results of the work.

  \section{Hyperfine structure in the system $c\bar s$, $c\bar u$, and $c\bar d$}
  Let us first consider the ground state of the $c\bar s$ system, corresponding to the $D_s(1968)$ meson and having the orbital angular momentum $L=0$ and spin $S_{tot}=0$, where $\bm S_{tot}=\bm S_{\bar s}+\bm S_c$. The first excited state, corresponding to the $D_s^*(2112)$ meson, has $L=0$ and $S_{tot}=1$ with a small admixture of a component having $L=2$, $S_{tot}=1$, and $J=1$, where $\bm J=\bm S_{tot}+\bm L$. Below we neglect this component, since it is unimportant for our problem. The difference in the masses of $D_s^*$ and $D_s$ is $\Omega=144\,\mbox{MeV}$. In the nonrelativistic approximation, the magnetic moment operator $\bm\mu$ of the system $c\bar s$ for $L=0$ is 
\begin{align} \label{munr}
	&\bm \mu=\left( \dfrac{e_{\bar s}}{m_s}\bm S_{\bar s}+\dfrac{e_{c}}{m_c}\bm S_{\bar s}\right)=\dfrac{1}{2}\left( \dfrac{e_{\bar s}}{m_s}+\dfrac{e_{c}}{m_c}\right)\bm S_{tot}
	+\dfrac{1}{2}\left( \dfrac{e_{\bar s}}{m_s}-\dfrac{e_{c}}{m_c}\right)\bm \Sigma,
\end{align}
where $m_s$ and $e_{\bar s}=e/3$ are the mass and charge of the $\bar s$ quark, $m_c$ and $e_{c}=2e/3$ are the mass and charge of the $c$ quark, $e$ is the charge of the proton, the operator $\bm \Sigma=\bm S_{\bar s}-\bm S_{c }$ has non-zero matrix elements between states with different values of $S_{tot}$, and the system of units $\hbar=c=1$ is used. The width $\Gamma(D_{s}^*\rightarrow D_s\gamma)$ in the nonrelativistic approximation is described by the usual formula for the $M1$ transition,
\begin{align} \label{Gamma0}
	&\Gamma_{nr}(D_{s}^*\rightarrow D_s\gamma)=\dfrac{4}{3}\Omega^3\,|t_0|^2\,,\quad t_0=\dfrac{1}{2}\left( \dfrac{e_{\bar s}}{m_s}-\dfrac{e_{c}}{m_c}\right).
\end{align}
This result is independent of the structure of the $c\bar s$ pair wave function. Substituting for the estimate
$m_s= 0.5\,\mbox{GeV}$ and $m_c= 1.7\,\mbox{GeV}\,$, we obtain the value $\Gamma_{nr}(D_{s}^*\rightarrow D_s\gamma)=0.547\,\mbox{KeV}\,$, which contradicts the experiment, $\Gamma(D_{s}^*\rightarrow D_s\gamma)\sim 0.114^{+0.066}_{-0.050}\,\mbox{KeV}\,$. The reason is that in the expression for $t_0$ there is a strong compensation between the contributions of the $\bar s$ and $c$ quarks, $t_0\approx e_{\bar s}/6m_s$. Therefore, it is necessary to take into account relativistic effects, which are important since the relative momentum of quarks is comparable to $m_s$. A similar effect was discussed in our recent work~\cite{Bondar:2025qzg} when considering the electric dipole transition $\Gamma(D_{s1}\rightarrow D_s\gamma)$, where $D_{s1}$ is the $c\bar s$ state of the system with $L=1$, $S_{tot}=1$ and $J=1$.
We will show that taking into account relativistic effects will lead to a strong decrease in the radiative width, compared to \eqref{Gamma0}, and to better agreement between the predictions and experimental data.

To derive the operator responsible for the hyperfine splitting, consider a particle with spin $1/2$, charge $e$, and mass $m$ in a magnetic field. We use the identity
$$(\bm\sigma\cdot\bm\pi)^2=\bm\pi^2 -e\,\bm\sigma\cdot\bm{\mathcal H}(\bm r,t),\quad \bm\pi=\bm p-e\bm A(\bm r,t)\,,$$
where $\bm\sigma$ are the Pauli matrices, $\bm p$ is the momentum operator, $\bm A(\bm r,t)$ is the vector potential, and $\bm{\mathcal H}(\bm r,t)$ is the corresponding magnetic field. Let us perform the expansion of the operator
$\sqrt{m^2+(\bm\sigma\cdot\bm\pi)^2}$ for small, compared to $m^2$, characteristic values of $\bm\pi^2$,
\begin{align}\label{exp0}
	&h=\sqrt{m^2+(\bm\sigma\cdot\bm\pi)^2}= m+\dfrac{(\bm\sigma\cdot\bm\pi)^2}{2m}-\dfrac{(\bm\sigma\cdot\bm\pi)^4}{8m^3}+\dots	\nonumber\\
	&=m + \dfrac{\bm\pi^2}{2m}-\dfrac{(\bm\pi^2)^2}{8m^3}    -\dfrac{e}{2m}\bm\sigma\cdot\bm{\mathcal H}+\dfrac{e}{8m^3}\{\bm\pi^2,\,\bm\sigma\cdot\bm{\mathcal H}\}+\dots\,,
\end{align} 
where $\{B,C\}=BC+CB$ and we have taken into account the terms linear in $\bm{\mathcal H}$.  Thus, we have obtained the Hamiltonian of the Pauli equation taking into account the first relativistic correction in $\bm\pi^2/m^2$. Let us now expand $h$ in the parameter $\bm\sigma\cdot\bm{\mathcal H}/(m^2+\bm\pi^2)$, assuming that the change in the magnetic field is small at $\delta r\lesssim 1/m$, but the characteristic values of $\bm\pi^2$ are not small compared to $m^2$. We have
 \begin{align}\label{exp1}
 	&h= \sqrt{m^2+\bm\pi^2}-\dfrac{e}{2}\left\{\dfrac{1}{\sqrt{m^2+\bm\pi^2}}\,,\,\bm s\cdot\bm{\mathcal H}\right\}+\dots\,,
  \end{align} 
  where $\bm s=\bm\sigma/2$ is the spin operator. To derive the contribution of the spin-spin interaction to the hyperfine splitting, we consider two particles with spin $1/2$, charges $e_1$ and $e_2$, and masses $m_1$ and $m_2$. To calculate the energy correction, we use the noncovariant perturbation theory,
  \begin{align}\label{PT}
 \Delta E=\sum_n\dfrac{|<n|\delta H|0>|^2}{E_0-E_n},
 \end {align}
 where $\delta H$ is the perturbation. Let $\Phi(\bm p)$ be the wave function of the bound state of the two-particle system in the momentum representation. There exists a spin-dependent energy correction $ \Delta E_S$ and a spin-independent correction $ \Delta E_0$.
  
 \subsection{Calculation of the correction $\Delta E_S$}
 It follows from Eqs.~\eqref{exp1} and \eqref{PT} that
   \begin{align}\label{hsS}
  	&\Delta E_S=2e_1e_2\iint \dfrac{d^3p\,d^3p'}{(2\pi)^6}\,\dfrac{\Phi^\dagger(\bm p') }{\sqrt{m_2^2+\bm p'^2}}\,\dfrac{4\pi}{2q}\,\dfrac{[\bm q\times \bm s_1]\cdot[\bm q\times \bm s_2]}{(-q)}\,\dfrac{\Phi(\bm p)}{\sqrt{m_1^2+\bm p^2}},
 \end{align}
 where $$\bm q=\bm p-\bm p'\,,\quad E-E_n=-q\,,\quad \bm{\mathcal H}_\lambda=-i\sqrt{\dfrac{4\pi}{2q}}\,[\bm q\times\bm e_\lambda]\,,\quad \sum_{\lambda=1,2} e_\lambda^a\,e_\lambda^b=\delta^{ab}-\dfrac{q^aq^b}{q^2}\,,$$ 
$\bm e_\lambda$ is the photon  polarization, $\lambda$ is its helicity, the factor $"2"$ in front of the integral takes into account the contribution of two equal amplitudes (the first particle emits and then the second absorbs and vice versa). We rewrite \eqref{hsS} as
  \begin{align}\label{hs1}
 	&\Delta E_S=\dfrac{4\pi g_{el}}{3}\iint \dfrac{d^3p\,d^3p'}{(2\pi)^6} \,\dfrac{\Phi^\dagger(\bm p')}{\sqrt{m_2^2+\bm p'^2}}\nonumber\\
 	&\times\left\{2\bm s_1\cdot\bm s_2+\left[\bm s_1\cdot\bm s_2-3\dfrac{(\bm s_1\cdot\bm q)(\bm s_2\cdot \bm q)}{q^2}\right]\right\}\,\dfrac{\Phi(\bm p)}{\sqrt{m_1^2+\bm p^2}},
 \end{align}
 where $g_{el}=-e_1e_2>0$. Using the identity
 \begin{align}\label{id}
& 2\bm s_1\cdot\bm s_2+\left[\bm s_1\cdot\bm s_2-3\dfrac{(\bm s_1\cdot\bm q)(\bm s_2\cdot \bm q)}{q^2}\right]\nonumber\\
&=\int d^3r\,\left\{2(\bm s_1\cdot\bm s_2)\,\delta(\bm r)-\dfrac{3}{4\pi}\Big[\bm (s_1\cdot\bm s_2)-3(\bm s_1\cdot\bm n)(\bm s_2\cdot \bm n)\Big]\right\}\,e^{-i\bm q\cdot\bm r},
\end{align}
where $\bm n=\bm r/r$, we can rewrite \eqref{hs1} in coordinate representation,
 \begin{align}\label{hs2}
	&\Delta E_S=g_{el}\int\!\! d^3r\,\Xi^\dagger(\bm r,m_2)\Big\{\dfrac{8\pi }{3}\delta(\bm r)(\bm s_1\cdot\bm s_2)\nonumber\\
	&+\dfrac{1}{3r^3}\Big[(\bm s_1\cdot\bm s_2)-3(\bm s_1\cdot\bm n)(\bm s_2\cdot \bm n)\Big]\Big\}\Xi(\bm r,m_1),\nonumber\\
	&\Xi(\bm r,m)=\int\dfrac{d^3p}{(2\pi)^3}\,\dfrac{\Phi(\bm p)}{\sqrt{m^2+\bm p^2}}\,e^{i\bm p\cdot\bm r}.
	\end{align}

  For $L=0$ we have
  \begin{align}\label{Om}
  		&\Delta E_S=\dfrac{4\pi g_{el}}{3}\, \Xi^\dagger(0,m_2)\,\left(\bm S_{tot}^2-\dfrac{3}{2}\right)\Xi(0,m_1).
\end{align}
From this formula we find an expression for the hyperfine splitting $\Omega_{el}$ in a system of interacting charges with different masses,
\begin{align}\label{Omf}
	&\Omega_{el}=\dfrac{8\pi g_{el}}{3}\, \Xi^\dagger(0,m_2)\Xi(0,m_1).
\end{align}

\subsection{Calculation of the correction $ \Delta E_0$}
The correction $\Delta E_0$ is the sum of two contributions, $\Delta E_0=\Delta E_{01}+\Delta E_{02}$. From the expansion
 \begin{align}\label{exp2}
&\sqrt{m^2+\bm\pi^2}=\sqrt{m^2+\bm p^2}  -\dfrac{e}{2}\left\{\dfrac{1}{\sqrt{m^2+\bm p^2}}\,,\,\bm A\cdot\bm p\right\}+\dots
\end{align} 
and Eq.~\eqref{PT}, we find similarly to \eqref{hsS}, 
\begin{align}\label{E01q}
&\Delta E_{01}=2g_{el}\iint \dfrac{d^3p\,d^3p'}{(2\pi)^6}\,\dfrac{\Phi^\dagger(\bm p') }{\sqrt{m_2^2+\bm p'^2}}\,\dfrac{4\pi}{2q}\,\dfrac{1}{(-q)}p'^{i}\left(\delta^{ij}-\dfrac{q^iq^j}{q^2}\right)p^{j}\,\dfrac{\Phi(\bm p)}{\sqrt{m_1^2+\bm p^2}},
\end{align}
where $\bm q=\bm p-\bm p'$. It is convenient to represent this formula as
\begin{align}\label{E01}
&\Delta E_{01}=-\dfrac{g_{el}}{2}\int\!\! d^3r\,\Xi^\dagger(\bm r,m_2)\,p^i\left(\dfrac{\delta^{ij}}{r}+\dfrac{r^ir^j}{r^3}\right)p^{j}\,\Xi(\bm r,m_1).
\end{align}
The  correction $\Delta E_{02}$ (the so-called Darwin term), although independent of spin, exists only for particles with spin. It is easy to obtain that
 \begin{align}\label{E02}
&\Delta E_{02}=\dfrac{\pi g_{el}}{2}\,\left[ \Xi^\dagger(0,m_1)\,\Xi(0,m_1)+\Xi^\dagger(0,m_2)\,\Xi(0,m_2)\right].
\end{align}

In our work we use a potential model in which the quark interaction potential $U(r)$ is described by the sum
 \begin{align}\label{pot}
 U(r)=U_{g }(r)+U_{conf }(r),\quad      U_{g }(r)=-\dfrac{g}{r},\quad U_{conf }(r)=br\,,
 \end{align}
 where $g$ and $b$ are the parameters of the model. The potential $U_{g}(r)$ is the zeroth component of the Lorentz vector, and $U_{conf}(r)$ is a Lorentz scalar~\cite{Godfrey:1985xj}. The energy corrections associated with $U_{g}(r)$ are obtained from the above formulas for electrodynamics by replacing $g_{el}\to g$. The potentials $U_{g}(r)$ and $U_{conf}(r)$ enter the quark–antiquark interaction Hamiltonian differently. In particular, there is a correction $\Delta E_{0b}$ to the spin-independent energy shift associated with the potential $U_{conf}(r)$. This correction is analogous to \eqref{E02} and has the form
   \begin{align}\label{E0b}
 &\Delta E_{0b}=\dfrac{b}{2}\,\int d^3r\,\Bigg[ \Xi^\dagger(\bm r,\,m_1)\, \left(\dfrac{1}{2r} -p^irp^i\right) \Xi(\bm r,\,m_1)\nonumber\\
 &+\Xi^\dagger(\bm r,\,m_2)\, \left(\dfrac{1}{2r} -p^irp^i\right) \Xi(\bm r,\,m_2)\Bigg].
 \end{align}
 The  correction $\Delta E_0$ is absent in the Izgur-Godfrey model~\cite{Godfrey:1985xj} but is consistent with the Breit Hamiltonian in quantum electrodynamics. The latter is obtained from the above formulas for the corrections by replacing $\sqrt{m^2+\bm p^2}$ with $m$ for both quarks. Note that $\Delta E_0$ does not contribute to the hyperfine splitting  but determines the overall level shift for states with $S_{tot}=0,\,1$.
 
 Since $U_{g}(r)$ is the zero component of the Lorentz vector, the hyperfine splitting in the $c\bar s$ system is given by Eq.~\eqref{Omf}, where $m_1=m_s$, $m_2=m_c$, and $g$ is the  parameter of potential.
 
 \subsection{Calculation of radiation widths}
 From Eq.~\eqref{exp1} it follows that $\Gamma(D_{s}^*\rightarrow D_s\gamma)$ in the leading approximation has the form
  \begin{align} \label{Gammarel}
 	&\Gamma(D_{s}^*\rightarrow D_s\gamma)=\dfrac{4}{3}\Omega^3\,|t_1|^2\,,\quad t_1=\dfrac{1}{2}\left( \dfrac{e_{\bar s}}{m_s}\,F_s-\dfrac{e_{c}}{m_c}\,F_c\right)\,,\nonumber\\
 	&F_s=m_s\int d^3r\,j_0(q_sr)\Xi^\dagger(r,m_s)\Psi(r)\,,\quad F_c=m_c\int d^3r\,j_0(q_cr)\Xi^\dagger(\bm r,m_c)\Psi(\bm r)\,,\nonumber\\
 	&q_s=\dfrac{M_{R}}{m_s}\Omega\,\quad q_c=\dfrac{M_R}{m_c}\Omega\,,\quad j_0(x)=\dfrac{\sin x}{x}\,,\quad M_R=\dfrac{m_sm_c}{m_s+m_c}.
 \end{align}
Note that for $L=0$ the functions $\Xi^\dagger(\bm r,m)$ and $\Psi(\bm r)$ are independent of the direction of  $\bm r$.

There is also a relativistic correction $\delta t_1$ to the amplitude $t_1$ in Eq.~ \eqref{Gammarel}, which follows from Eq.~(21) of our paper~\cite{Bondar:2025qzg} and is related to the influence of the quark-antiquark interaction on the radiation process. The 
corrections $\delta F_s$ and $\delta F_c$ to the corresponding form factors in \eqref{Gammarel}, following from the amplitude $\delta t_1$, read
\begin{align}\label{delFsc}
	&\delta F_s= -M_R\int d^3r\, \dfrac{j_1(q_sr)}{q_sr}\Xi^\dagger(r,m_s)\Bigg[\dfrac{1}{2}\left(br-\dfrac{g}{r}\right)
	\Xi(r,m_s)+\dfrac{g}{r}\Xi(r,m_c)\Bigg]\,,\nonumber\\
	&\delta F_c= -M_R\int d^3r\, \dfrac{j_1(q_cr)}{q_cr}\Xi^\dagger(r,m_c)\Bigg[\dfrac{1}{2}\left(br-\dfrac{g}{r}\right)
	\Xi(r,m_c)+\dfrac{g}{r}\Xi(r,m_s)\Bigg]\,,\nonumber\\
	&j_1(x)=\dfrac{\sin x}{x^2}-\dfrac{\cos x}{x}\,.
\end{align}
If we neglect the retardation effect and tend $q_s\to 0$ and $q_c\to 0$, then we get \eqref{delFsc} with the replacement $j_1(x)/x\to 1/3$.
 
 In order to obtain predictions for the hyperfine splitting and the transition width, it is necessary to know the wave function, which we will discuss in the next section.
 
 \section{Wave function}
 The wave function in the leading approximation in our model satisfies the equation 
  \begin{align} \label{WF}
  	&E\Phi(\bm p)=\left[\sqrt{m_s^2+\bm p^2}+\sqrt{m_c^2+\bm p^2}-\dfrac{g}{r}+br\right]\Phi(\bm p)\,.
  \end{align}
Here $r$ and $1/r$ are integral operators. It is convenient to solve this equation using the variational principle. For $L=0$ we represent $\Phi(\bm p)$ as
 \begin{align} \label{WF1}
 	&\Phi(\bm p)=\sqrt{\dfrac{N}{4\pi}}\,\chi\,\exp\left[-\dfrac{p^2}{2\omega_0^2}\right]\,,\quad N=\dfrac{32\pi^{5/2}}{\omega_0^3}\,,
 \end{align}
where $\chi$ is the spin part of the wave function and $\omega_0$ is the variational parameter. The corresponding wave function in the coordinate representation is 
 \begin{align} \label{WF11}
 	&\Psi(\bm r)=\sqrt{\dfrac{{\cal N}}{4\pi}}\,\chi\,\exp\left[-\dfrac{\omega_0^2 r^2}{2}\right]\,,\quad {\cal N}=\dfrac{4\omega_0^3}{\sqrt{\pi}}\,.
 \end{align}
 For such wave functions, the average value $E_0$ of the Hamiltonian in Eq.~\eqref{WF} is
  \begin{align} \label{Evar}
  		&E_0=m_s{\cal F}_0\left(\dfrac{m_s^2}{2\omega_0^2}\right)+m_c{\cal F}_0\left(\dfrac{m_c^2}{2\omega_0^2}\right)+
  		\dfrac{2}{\sqrt{\pi}}\left(\dfrac{b}{\omega_0}-g\omega_0\right)\,,\nonumber\\
  		&{\cal F}_0(x)=\sqrt{\dfrac{2x}{\pi}}\,e^x\, K_1(x),
 \end{align}
where $K_n(x)$ is the  modified Bessel function of the third kind. The variational parameter $\omega_0$ is found from the equation $\partial E_0/\partial\omega_0=0$ or
 \begin{align} \label{dis}
&\dfrac{m_s}{\omega_0}\left[{\cal F}_0\left(\dfrac{m_s^2}{2\omega_0^2}\right)-{\cal F}_1\left(\dfrac{m_s^2}{2\omega_0^2}\right)\right]+\dfrac{m_c}{\omega_0}\left[{\cal F}_0\left(\dfrac{m_c^2}{2\omega_0^2}\right)-{\cal F}_1\left(\dfrac{m_c^2}{2\omega_0^2}\right)\right]\nonumber\\
&-\dfrac{2}{\sqrt{\pi}}\left(\dfrac{b}{\omega_0^2}+g\right)=0\,,\nonumber\\
&{\cal F}_1(x)=\sqrt{\dfrac{(2x)^3}{\pi}}\,e^x\,[ K_1(x)-K_0(x)].
\end{align}
Note that ${\cal F}_0(x)\to 1$, ${\cal F}_1(x)\to 1$, and $ {\cal F}_0(x)-{\cal F}_1(x)\to 3/(4x) $ for $x\to\infty$. For $x\to 0$ we have ${\cal F}_0(x)\to \sqrt{2/(\pi x)}$ and ${\cal F}_1(x)\to \sqrt{8x/\pi}$.

It is interesting that Eq.~\eqref{dis} in the limit $m_s\to 0$ and $m_c\to \infty$  has no singularities and its solution and the corresponding energy $E_0$ for $g<1$ are
\begin{align} \omega_0=\sqrt{\dfrac{b}{1-g}}\,,\quad E_0=m_c+4\,\sqrt{\dfrac{(1-g)\,b}{\pi}}\,.
\end{align}
Here the singularity for $g>1$ is similar to that which exists in the solution of the Dirac equation in the Coulomb field. This singularity is absent for a large but finite mass $m_c$.

\section{Hyperfine splitting and form factors}
Using variational functions, for the function $\Xi(\bm r,m)$ defined in Eq.~\eqref{hs1}, we obtain
	\begin{align}\label{Xi}
		&\Xi(r,m)=\dfrac{1}{m}\sqrt{\dfrac{{\cal N}}{4\pi}}\,\dfrac{2}{\sqrt{\pi}}\int_0^\infty \dfrac{ds}{(1+2\omega_0^2s^2/m^2)^{3/2}}\,
		\exp\left[-s^2-\dfrac{\omega_0^2r^2}{2(1+2\omega_0^2s^2/m^2)}\right].
		\end{align} 
	Then, for the hyperfine splitting $\Omega$ \eqref{Omf} we find
\begin{align}\label{Omfvar}
	&\Omega=\dfrac{8g\,\omega_0^3}{3\sqrt{\pi}\,m_sm_c}\,{\cal F}_1\left(\dfrac{m_s^2}{4\omega_0^2}\right){\cal F}_1\left(\dfrac{m_c^2}{4\omega_0^2}\right),
\end{align} 
where the function ${\cal F}_1$ is defined in \eqref{dis}. In the limit of small $m_s$ and large $m_c$, the hyperfine splitting $\Omega$ is 
\begin{align}\label{Omas}
	&\Omega=\dfrac{8\sqrt{2}g\,b}{3\pi\,(1-g)\, m_c}.
\end{align} 
It is seen that the asymptotics of $\Omega$ is independent of $m_s$.

The spin-independent correction $\Delta E_0$ to the energy level is 
\begin{align} \label{DE0}
&\Delta E_0=\Delta E_{01}+\Delta E_{02}+\Delta E_{0b}\,,\nonumber\\
&\Delta E_{01}=	-\dfrac{g\omega_0}{\pi^{3/2}}\int_0^\infty dy\,e^{-y}\,
e^{z_s+z_c}K_0(z_s)K_0(z_c),\nonumber\\ 
&\Delta E_{02}=\dfrac{g\,\omega_0^3}{2\sqrt{\pi}}\left[\dfrac{1}{\,m_s^2}{\cal F}_1^2\left(\dfrac{m_s^2}{4\omega_0^2}\right)+\dfrac{1}{\,m_c^2}{\cal F}_1^2\left(\dfrac{m_c^2}{4\omega_0^2}\right)\right],\nonumber\\
&\Delta E_{0b}= \dfrac{b}{8\omega_0\pi^{3/2}}\int_0^\infty dy\,e^{-y}\,\Bigg\{e^{2z_s}\Bigg[2K_0^2(z_s)-y^2[K_0(z_s)+K_1(z_s)]^2\Bigg]\nonumber\\
&+e^{2z_c}\Bigg[2K_0^2(z_c)-y^2[K_0(z_c)+K_1(z_c)]^2\Bigg]	\Bigg\},\nonumber\\
&	z_s=\dfrac{1}{4}\left(y+\dfrac{m_s^2}{\omega_0^2}\right)\,,\quad	z_c=\dfrac{1}{4}\left(y+\dfrac{m_c^2}{\omega_0^2}\right).
\end{align}
Finally, for energy levels with $S_{tot}=0$ and $S_{tot}=1$ we have
\begin{align} \label{Etot}
&E(S=0)=E_0+\Delta E_0 -\dfrac{3}{4}\Omega\,,\quad E(S=1)=E_0+\Delta E_0 +\dfrac{1}{4}\Omega .
\end{align}

For the form factors $F_s$ and $F_c$ in Eq.~\eqref{Gammarel} we obtain
 \begin{align}\label{FsFcret}
	&F_s=\dfrac{2m_s}{\sqrt{\pi}\omega_0}\int_0^\infty \dfrac{ds}{(s^2+1)^{3/2}}\,
	\exp\left[-\dfrac{m_s^2}{\omega_0^2}s^2-Q_s^2\dfrac{2s^2+1}{4(s^2+1)}\right]\,,\nonumber\\
	&\quad F_c=\dfrac{2m_c}{\sqrt{\pi}\omega_0}\int_0^\infty \dfrac{ds}{(s^2+1)^{3/2}}\,
	\exp\left[-\dfrac{m_c^2}{\omega_0^2}s^2-Q_c^2\dfrac{2s^2+1}{4(s^2+1)}\right]\,,\nonumber\\
	&Q_s=\dfrac{M_R\Omega}{m_s\omega_0}\,,\quad Q_c=\dfrac{M_R\Omega}{m_c\omega_0}\,.
\end{align} 
In the limit $Q_s\to 0$ and $Q_c\to 0$ (that is, without the retardation effect taken into account) we have
 \begin{align}\label{FsFc}
 	&F_s={\cal F}_1\left(\dfrac{m_s^2}{2\omega_0^2}\right)\,,\quad F_c={\cal F}_1\left(\dfrac{m_c^2}{2\omega_0^2}\right).
 \end{align} 

 Let us now discuss the corrections $\delta F_s$ and $\delta F_c$ to the form factors defined in \eqref{delFsc}. Using the representation \eqref{Xi}, we obtain the corrections without the retardation effect taken into account.
   \begin{align}\label{delFsc1}
  	&\delta F_s= -\dfrac{M_R}{3\omega_0\pi^{3/2}}\int_0^\infty dy\,e^{-y}\,\Bigg\{\dfrac{1}{2}e^{2z_s}\left[-gK_0^2(z_s)+\dfrac{by}{4\omega_0^2}[K_0(z_s)+K_1(z_s)]^2\right]\nonumber\\
    	  	&+ge^{z_s+z_c}K_0(z_s)K_0(z_c)  	\Bigg\}\,,\nonumber\\ 
  	  		&\delta F_c= -\dfrac{M_R}{3\omega_0\pi^{3/2}}\int_0^\infty dy\,e^{-y}\,\Bigg\{ \dfrac{1}{2}e^{2z_c}\left[-gK_0^2(z_c)+\dfrac{by}{4\omega_0^2}[K_0(z_c)+K_1(z_c)]^2\right]\nonumber\\
  	   	  	&+ge^{z_s+z_c}K_0(z_s)K_0(z_c)  	\Bigg\}\,,\nonumber\\ 
  	  &	z_s=\dfrac{1}{4}\left(y+\dfrac{m_s^2}{\omega_0^2}\right),\quad	z_c=\dfrac{1}{4}\left(y+\dfrac{m_c^2}{\omega_0^2}\right).
  \end{align} 
  Taking the retardation effect into account, we obtain for the correction $\delta F_s^{(ret)}$,
   \begin{align}\label{delFsret}
  	&\delta F_s^{(ret)}= -\dfrac{8M_R}{\omega_0\pi^{3/2}}\iint_0^\infty dt_1dt_2\,\dfrac{\exp(-t_1^2/2-t_2^2/2)}{\sqrt{t_1^2+m_s^2/\omega_0^2}} \nonumber\\
  	&\times\Bigg[g\left(-\dfrac{1}{2\sqrt{t_2^2+m_s^2/\omega_0^2}}+\dfrac{1}{\sqrt{t_2^2+m_c^2/\omega_0^2}}\right)\,G_1\left(\dfrac{t_1}{Q_s},\dfrac{t_2}{Q_s}\right)\nonumber\\
  	&+\dfrac{b}{2\omega_0^2Q_s^2\sqrt{t_2^2+m_s^2/\omega_0^2}}\,G_2\left(\dfrac{t_1}{Q_s},\dfrac{t_2}{Q_s}\right)\Bigg]\,.
  \end{align} 
  The following functions are used here
 \begin{align}\label{G1G2}
	&G_1(x,y)=f_1(x,y)+f_1(-x,-y)-f_1(x,-y) -f_1(-x,y)+\dfrac{1}{3}xy\,,\nonumber\\
&G_2(x,y)=f_2(x,y)+f_2(-x,-y)-f_2(x,-y) -f_2(-x,y)\,,\nonumber\\
	&f_1(x,y)=\dfrac{1}{48}(2-x-y)(1+x+y)^2\ln(1+x+y)^2\,,\nonumber\\
	&f_2(x,y)=\dfrac{1}{8}(x+y)\ln(1+x+y)^2\,.
\end{align}   
The correction $\delta F_c^{(ret)}$ is obtained from $\delta F_s^{(ret)}$ by replacing $m_s\leftrightarrow m_c$ and $Q_s\to Q_c$. For all reasonable values of the parameters, the corrections $\delta F_s^{(ret)}$ and $\delta F_c^{(ret)}$ from Eq.~\eqref{delFsret} differ slightly from $\delta F_s$ and $\delta F_c$ from Eq.~\eqref{delFsc1}.

\section{Parameters of the model and numerical results}

Using  obtained expressions for $E(S=0)$ and $\Omega$, we fix the parameters of the model. In total, we have five parameters: the mass of the light quark $m_l$, the mass of the strange quark $m_s$ and the mass of the heavy quark $m_Q$ for $D$ mesons and $B$ mesons, the dimensionless parameter $g$ proportional to $\alpha_s$, and the parameter $b$ characterizing the confinement potential. For a pre-selected value of $m_l$ in the range from 20 to 300$\,\mbox{MeV}$, we choose the values of the parameters $m_s$, $m_Q$, $g$, and $b$ so that the values of $E(S=0)$ and $\Omega$ correspond to the experimental values for the masses $D^+$,$D^{*+}$,$D^+_s$, and $D^{*+}_s$. In this case, the parameters $m_Q$, $g$ and $b$ are universal for all four mesons. Examples of the values of the resulting parameters and the corresponding meson masses in our model are given in Table~\ref{tb1}.

\begin{table}[!h]
		\caption{\label{tb1} The values of the  parameters in our model
		}
		\begin{center}
			
			%\begin{ruledtabular}
			\begin{tabular}{lrrrrrrrrrr} 
				& \multicolumn{1}{c}{ }&\multicolumn{1}{c}{ }&\multicolumn{1}{c}{ }&\multicolumn{1}{c}{ }&\multicolumn{1}{c}{ }&\multicolumn{1}{c}{ }&\multicolumn{1}{c}{ }&\multicolumn{1}{c}{ }&\multicolumn{1}{c}{ }\\

				Parameters & \multicolumn{1}{c}{ $m_l$ (MeV)}  &\multicolumn{1}{c}{ $m_Q$ (MeV) } & \multicolumn{1}{c}{ $g$   }  &\multicolumn{1}{c}{ $b$ ($GeV^2$)  } &\multicolumn{1}{c}{}  &\multicolumn{1}{c}{  } & \multicolumn{1}{c}{ $M(D_{(s)})$ (MeV)  }&\multicolumn{1}{c}{ $M(D_{(s)}^{*})$ (MeV) } &\multicolumn{1}{c}{ } &\multicolumn{1}{c}{ }\\
				\hline
				
				$ c\bar{d} $ I  & 20 & 1805 &\,\, 0.934 &\,\, 0.100 &  &  & 1870\,\,\,\,\,  & 2010\,\,\,\,\, &  &\\
				$ c\bar{s} $ I & 270 & 1805 & 0.934 & 0.100 &  &  & 1971\,\,\,\,\, & 2113\,\,\,\,\, &  &\\
                    $ c\bar{d}$ II & 133 & 1749 & 0.91 & 0.110 &  &  & 1869\,\,\,\,\, & 2009\,\,\,\,\, &  &\\
                    $ c\bar{s} $ II & 320 & 1749 & 0.91  & 0.110 &  &  & 1970\,\,\,\,\, & 2113\,\,\,\,\,  &  &\\
                    	$ c\bar{d}$ III & 300 & 1646 & 0.896  & 0.110 &  &  & 1869\,\,\,\,\,  & 2009\,\,\,\,\, &  &\\ 
				$ c\bar{s}$ III & 454 & 1646 & 0.896  & 0.110 &  &  & 1969\,\,\,\,\, & 2112\,\,\,\,\,  &  &\\
                    $ c\bar{d} $ Experiment&  &  &   &  &  &  & 1869.5 $\pm$ 0.4 & 2010.26 $\pm$ 0.05 &  &\\
                    $ c\bar{s} $ Experiment&  &  &   &  &  &  & 1968.35 $\pm$ 0.4 & 2112.2 $\pm$ 0.4  &  &\\
				\hline
				
			\end{tabular}
			%\end{ruledtabular}
			
		\end{center}
	\end{table}
For the  parameters of the model fixed in this way, we calculate the widths of the radiative transitions, the results are given in Table~\ref{tb2}.

\begin{table}[!h]
		\caption{\label{tb2} The values of the obtained radiative widths in $\mbox{KeV}$.
		}
		\begin{center}
			
			%\begin{ruledtabular}
			\begin{tabular}{lrrrrrrrrrr} 
				& \multicolumn{1}{c}{ }&\multicolumn{1}{c}{ }&\multicolumn{1}{c}{ }&\multicolumn{1}{c}{ }&\multicolumn{1}{c}{ }&\multicolumn{1}{c}{ }&\multicolumn{1}{c}{ }&\multicolumn{1}{c}{ }&\multicolumn{1}{c}{ }\\

				Parameters & \multicolumn{1}{c}{ }  &\multicolumn{1}{c}{ $\Gamma(D^{*0})$  } & \multicolumn{1}{c}{    }  &\multicolumn{1}{c}{ $\Gamma(D^{*+})$    } &\multicolumn{1}{c}{}  &\multicolumn{1}{c}{ $\Gamma(D^{*+}_s)$  } & \multicolumn{1}{c}{   }&\multicolumn{1}{c}{  } &\multicolumn{1}{c}{ } &\multicolumn{1}{c}{ }\\
				\hline
				
				 I  &  &  19.0\ \ \ \ \ \ &  & 0.570\ \ \ \ \ \  &  & 0.250\ \ \ \ \  &   &  &  &\\
			
                II &  & 17.0\ \ \ \ \ \  &  & 0.430\ \ \ \ \ \  &  & 0.150\ \ \ \ \  &  &  &  &\\
                
                 III &   & 13.0\ \ \ \ \ \  &  & 0.140\ \ \ \ \ \  &  & 0.030\ \ \ \ \  &  &  &  &\\ 
                
                Experiment&  & 20.2 $\pm$ 0.45  &  & 1.33 $\pm$ 0.36 &  & $0.114^{+0.066}_{-0.050}$  &  &  &  &\\
                
				\hline
				
			\end{tabular}
		\end{center}
	\end{table}

From the comparison of the obtained values for the radiative widths, we can conclude that the best agreement with the experiment in our model is achieved at the minimum value of the light quark mass. It is interesting to note that the model is stable for $m_l$ tending to zero, which was expected from the asymptotics of \eqref{Omas}. Note, however, that the radiative width for $D^{*+}$ predicted by us for all sets of model parameters is noticeably lower than the result of the CLEO collaboration~\cite{CLEO:1997rew}  obtained many years ago. Therefore, it would be desirable to repeat this measurement at a qualitatively new level of accuracy, which is possible on the BELLE-II and BES-III detectors.

We can apply our model to $B$ mesons, assuming that the masses of the light quarks $m_l$, $m_s$, and the parameter $b$ remain the same as for $D$ mesons, and the parameters $g$ and $m_Q$ are chosen so as to reproduce the experimental values of the $B^0$ and $B^{*0}$ meson masses. In this case, we are already able to predict the $B_s$ mass and the hyperfine splitting of the $B_s$ and $B_s^*$ meson masses (Table~\ref{tb3}). At the time of writing, at the EPS-HEP25 conference, the CMS collaboration presented new, more precise measurements of the hyperfine splitting of the $B_{(s)}$ meson masses~\cite{CMS:2025kat}.  We  compare the results of our calculations with the result of \cite{CMS:2025kat}.

 \begin{table}[!h]
		\caption{\label{tb3} Values of the obtained masses and hyperfine splitting values for $B$ mesons in $\mbox{MeV}$.
		}
		\begin{center}
			
			%\begin{ruledtabular}
			\begin{tabular}{lrrrrrrrrrr} 
				& \multicolumn{1}{c}{ }&\multicolumn{1}{c}{ }&\multicolumn{1}{c}{ }&\multicolumn{1}{c}{ }&\multicolumn{1}{c}{ }&\multicolumn{1}{c}{ }&\multicolumn{1}{c}{ }&\multicolumn{1}{c}{ }&\multicolumn{1}{c}{ }\\

				Parameters & \multicolumn{1}{c}{$M(B^{0})$ }  &\multicolumn{1}{c}{ $M(B^{*0})$  } & \multicolumn{1}{c}{$M(B^{0}_s)$    }  &\multicolumn{1}{c}{$M(B^{*0}_s)$    } &\multicolumn{1}{c}{$\Delta M (B^{*0})$}  &\multicolumn{1}{c}{ $\Delta M (B^{*0}_s)$  } & \multicolumn{1}{c}{  }&\multicolumn{1}{c}{  } &\multicolumn{1}{c}{ } &\multicolumn{1}{c}{ }\\
				\hline
				
				 I  & 5281\ \ \ \ \ \ &  5326\ \ \ \ \ \ & 5377\ \ \ \ \ \   & 5425\ \ \ \ \ \   & 45.0\ \ \ \ \ \ \ & 48.1\ \ \ \ \ \ \   &   &  &  &\\
			
                II  & 5279\ \ \ \ \ \ &  5325\ \ \ \ \ \ & 5376\ \ \ \ \ \   & 5425\ \ \ \ \ \   & 45.5\ \ \ \ \ \ \ & 48.6\ \ \ \ \ \ \   &   &  &  &\\
                
                 III & 5280\ \ \ \ \ \   & 5326\ \ \ \ \ \   & 5376\ \ \ \ \ \  & 5425\ \ \ \ \ \   & 46.0\ \ \ \ \ \ \  & 49.2\ \ \ \ \ \ \   &  &  &  &\\ 
                
                Experiment& 5279.6 $\pm$ 0.2 & 5325.2 $\pm$ 0.2  & 5366.9 $\pm$ 0.11  & 5416.3 $\pm$ 0.2 & 45.47 $\pm$ 0.07 & 49.41 $\pm$ 0.15  &  &  &  &\\
                
				\hline
				
			\end{tabular}
			\end{center}
	\end{table}

Comparison of the results obtained in the model with the experiment shows that the masses of the $B_s$ and $B_s^*$ mesons are overestimated by 9-10 $\mbox{MeV}$, which seems quite natural, since the effective mass of a quark is not a constant and depends on the size of the quark wave function in the meson. Since the reduced mass of the $s$ quark for the $B$ meson is larger than for the $D$ meson, one can expect that the effective mass will be slightly smaller. At the same time, the hyperfine splitting depends much less on the mass of the light quark and is in remarkable agreement with the experimental value. We believe that this success of the model is completely nontrivial. As is known, in the heavy quark approximation in QCD, mass relations were obtained that relate the hyperfine splitting values for $D^*_{(s)}$ and $B^*_{(s)}$~mesons~\cite{Rosner:1992qw,Goity:2007fu,Karliner:2019lau}:
      \begin{align}\label{delta_m}
 	&\dfrac{(M(B^{*0}_s) - M(B^{0}_s)) - (M(B^{*0}) - M(B^{0}))}{(M(D^{*+}_s) - M(D^{+}_s)) - (M(D^{*+}) - M(D^{+}))}  = \dfrac{m_c\chi(m_b)}{m_b\chi(m_c)},\\
  &\nonumber\\
   	&\dfrac{((M(B^{*+}) - M(B^{+}))+2(M(B^{*0}_s) - M(B^{0}_s)))}{((M(D^{*0}) - M(D^{0}))+2(M(D^{*+}_s) - M(D^{+}_s))}=\dfrac{m_c\chi(m_b)}{m_b\chi(m_c)} ,
  \end{align}
    where $m_c$ and $m_b$ are the masses of the $c$ and $b$ quarks, and $\chi(m_Q)$ is a slowly varying function of the arguments. The above relations are obtained up to expansion terms of the order of ${\cal{O}}(1/m_Q^3)$,${\cal{O}}(\alpha /m_Q^2$),${\cal{O}}(m_l^{3/2})$ and ${\cal{O}}(m_l/m_Q^2)$~\cite{Goity:2007fu}. According to these relations, the difference in the hyperfine splittings for $B_s$ and $B^0$ mesons is expected to be approximately three times smaller than the corresponding difference for $D$ mesons.
  Now, thanks to the CMS measurements, it can be seen that these relations cannot be simultaneously reconciled with experiment:
     \begin{align}\label{delta_m_exp}
 	&\dfrac{(2.936 \pm 0.15 MeV)}{( 3.2 \pm 0.4 MeV)}  = \dfrac{m_c\chi(m_b)}{m_b\chi(m_c)},
 	&\dfrac{(144.1 \pm 0.28 MeV)}{( 429.6 \pm 0.8 MeV)}  = \dfrac{m_c\chi(m_b)}{m_b\chi(m_c)}.
  \end{align} 

Our model reproduces the difference in hyperfine splittings for $B$ mesons with an accuracy of 0.5~$\mbox{MeV}$ regardless of the choice of $m_l$.

\begin{table}[ht]
	\caption{Radiative widths of $\Gamma(B^*_{(s)}\to B_{(s)}\gamma)$ decays in $\mbox{KeV}$. The range of values  is defined as the variation of the calculated width for sets I-III of the parameters.}
	\label{tab:width_VPg_B}
	\begin{center}
		\begin{tabular}{c|c|c|c}
			\hline\hline
			Ref.  & $\Gamma(B^{*+}\to B^+\gamma)$  &  $\Gamma(B^{*0}\to B^0\gamma)$ & $\Gamma(B^{*+}_s\to B^+_s\gamma)$\\
			\hline
			BSE~\cite{Bhatnagar:2020vpd}   & $--$ & 0.136 & 0.062\\
			RQM~\cite{Ebert:2002xz} & 0.19 & 0.070 & 0.054\\		
			\cite{Godfrey:2016nwn} & 4.31 & 1.23 & 0.313\\
			\cite{Lu:2016bbk} & $--$ & 0.10 & 0.10 \\
			\cite{Kher:2017mky} & $--$ & 0.069 & 0.095\\				
			\cite{Patel:2022hhl} & $--$ & 0.135 & 0.102   \\
			This work & 0.21 - 0.31  & 0.076 - 0.106 &  0.068 - 0.098 \\
			\hline\hline
		\end{tabular}
	\end{center}
\end{table}

It is useful to compare the hyperfine splitting ratio for $B^{*0}$ and $B^{*0}_s$ obtained in our model with experiment~\cite{CMS:2025kat} and with the results of lattice calculations~\cite{Dowdall:2012ab}:
\begin{align}\label{Hyper}
 	&R = \dfrac{(M(B^{*0}) - M(B^{0}))}{(M(B^{*0}_s) - M(B^{0}_s))},\nonumber\\
 	&R_{this \  work} = 0.935(10),\  R_{lat} = 0.993(33)(5),\  R_{exp} = 0.920(3)(0.7). 
  \end{align} 

In Table~\ref{tab:width_VPg_B} we compare our calculated values for the widths of the radiative $\mbox{M1}$ transitions of $B^*_{(s)}$ mesons  with the results of other authors. It is worth noting that in our model the sensitivity of the  $B^*_{(s)}$ meson  radiative decay widths to the  parameters of the model is significantly smaller than in the case of $D^*_{(s)}$ mesons.

 \section{Conclusion}
 In our work, we propose a relativistic generalization of the potential quark model. This model allows us to calculate the radiative widths of heavy-light mesons and predict the hyperfine splitting value in the $b\bar s$ quark system. The obtained hyperfine splitting value is in perfect agreement with the results of the CMS~\cite{CMS:2025kat} detector.

\bibliographystyle{JHEP}
\bibliography{note}

\end{document}